\documentclass[
]{ceurart}

\usepackage{listings}
\newcommand\boldparagraph[1]{\vspace{0.35em}\noindent\textbf{#1}}
\newcommand{\rowgroup}[1]{\hspace{-1em}#1}

\newcommand{\doctfquery}{\textsf{DocT5Query}}

\newcommand{\deeperimpact}{\textsf{DeeperImpact}}

\newcommand{\msmarcodev}{\textsf{MS MARCO Dev Queries}\xspace}
\newcommand{\trecdl}{\textsf{TREC 2019}\xspace}
\newcommand{\trecdltw}{\textsf{TREC 2020}\xspace}

\newcommand{\multibert}{\textsf{DeepImpact}}
\newcommand{\splade}{\textsf{SPLADE}}
\newcommand{\unicoil}{\textsf{uniCOIL}}
\newcommand{\dtqmm}{Doc2Query-\textit{-}}

\begin{document}

\copyrightyear{2024}
\copyrightclause{Copyright for this paper by its authors. Use permitted under Creative Commons License Attribution 4.0 International (CC BY 4.0).}
\conference{ReNeuIR 2024 (at SIGIR 2024) -- 3rd Workshop on Reaching Efficiency in Neural Information Retrieval, 18 July, 2024, Washington D.C, USA}

\title{DeeperImpact: Optimizing Sparse Learned Index Structures}


\author[1]{Soyuj Basnet}[%
email=soyuj@nyu.edu,
]

\address[1]{NYU Abu Dhabi, UAE}
\address[2]{New York University, US}
\address[3]{Pinecone, US}

\author[2]{Jinrui Gou}[%
email=jg6226@nyu.edu,
]

\author[3]{Antonio Mallia}[%
email=antonio@pinecone.io,
]

\author[2]{Torsten Suel}[%
email=torsten.suel@nyu.edu,
]


\begin{abstract}
A lot of recent work has focused on sparse learned indexes that use deep neural architectures to significantly improve retrieval quality while keeping the efficiency benefits of the inverted index. While such sparse learned structures achieve effectiveness far beyond those of traditional inverted index-based rankers, there is still a gap in effectiveness to the best dense retrievers, or even to sparse methods that leverage more expensive optimizations such as query expansion and query term weighting.

We focus on narrowing this gap by revisiting and optimizing \multibert{}, a sparse retrieval approach that uses DocT5Query for document expansion followed by a BERT language model to learn impact scores for document terms. We first reinvestigate the expansion process and find that the recently proposed \dtqmm{} query filtration does not enhance retrieval quality when used with \multibert{}. Instead, substituting T5 with a fine-tuned Llama 2 model for query prediction results in a considerable improvement. Subsequently, we study training strategies that have proven effective for other models, in particular the use of hard negatives, distillation, and pre-trained CoCondenser model initialization. Our results substantially narrow the effectiveness gap with the most effective versions of SPLADE.

\end{abstract}

\begin{keywords}
  learned index \sep
  sparse representations \sep
  document expansion \sep
  large language models 
\end{keywords}

\maketitle

\section{Introduction}

Traditional IR systems rely heavily on complex textual matching algorithms and frequency-based ranking functions such as BM25 for fast retrieval. Despite being highly computationally efficient, these techniques suffer from the vocabulary mismatch problem when semantically relevant queries do not syntactically match the keywords indexed for the documents. With the emergence of the transformer architecture \cite{vaswani2023attention} and large pre-trained language models such as BERT \cite{devlin2018bert}, neural models have been able to leverage the context of the queries and documents to achieve impressive state-of-the-art retrieval quality \cite{nogueira2019bertranker}. Unfortunately, this comes at the cost of expensive query processing times, making them slow, difficult to scale, and also requiring new infrastructures to deploy. Thus, there has been an interest in developing learned sparse solutions that combine the efficiency of the traditional inverted index architecture \cite{BMP, mmmp22-emnlp, mallia2022faster} with the retrieval quality of neural models.

\subsection{Related Works}

Early in the development of sparse methods, \citet{deepct} introduced DeepCT, which uses contextual word representations from BERT to generate more effective document term weights that can be stored and used in inverted indexes. DeepCT estimates a term's context-specific score by learning to map its final BERT hidden state to a single term weight. However, one of the main limitations of DeepCT is that it still suffers from the vocabulary mismatch problem, as it only assigns weights to terms present in the document. Furthermore, as highlighted by \citet{mallia2021learning}, DeepCT is trained as a \textit{per-token regression task}, which requires knowing the ground truth for each term weight, and thereby prevents individual scores to co-adapt for identifying relevant documents.

In a separate approach, \citet{docTTTTTquery} proposed a document expansion strategy called \doctfquery{} that trains T5 \cite{tfive}, a sequence-to-sequence model, to predict relevant queries that a given document might potentially answer. By appending these predicted queries as part of the document before indexing, they achieved enhanced retrieval metrics using BM25 scores, and attributed these improvements to the \textit{injection} of new terms as well as \textit{reweighing} of existing terms. Alternatively, \citet{bai2020sparterm} introduced SparTerm, which provides another way of performing document expansion---by predicting an importance score for all terms in the vocabulary and using a gating mechanism to keep only a sparse subset. Finally, this sparse subset is used to learn an \textit{end-to-end} score for relevant and non-relevant documents.

\multibert{} \cite{mallia2021learning} made improvements by building on the key ideas of DeepCT and \doctfquery{}. It first uses \doctfquery{} to expand the documents in the dataset with potentially relevant query terms in an attempt to solve the vocabulary mismatch problem. Then, instead of learning independent term-level scores like DeepCT, \multibert{} is trained to directly optimize the sum of query term impacts such that it maximizes the score difference between the relevant and non-relevant passages for a given query.

In another important line of work, \citet{formal2021splade} introduced SPLADE, a sparse retrieval model that improves upon the key ideas of SparTerm by adding a log-saturation effect and changes to the training mechanism. It predicts the term importance for all the tokens in the BERT WordPiece Tokenizer vocabulary. SPLADE has been continuously optimized over the last few years, most recently with the release of SPLADE-v3 \cite{spladev3}, which obtains significant improvements in quality over earlier versions.

Other important approaches for sparse learned indexing include COIL \cite{gao2021coil}, the uniCOIL variant by \citet{lin2021brief}, and TILDE \cite{zhuang2021tilde, zhuang2021fastpassagererankingcontextualized}. COIL \cite{gao2021coil} uses a deep language model to encode query and document tokens into contextualized vectors, stored directly in the inverted index. The retrieval process involves calculating inner products between matching token vectors, combining the efficiency of exact lexical match with the enhanced representational capabilities of LLMs. uniCOIL \cite{lin2021brief} is a simplified variant of COIL that reduces token dimensions to scalar weights, making it directly comparable to DeepCT and DeepImpact. When combined with document expansion techniques like \doctfquery{}, uniCOIL achieves effectiveness comparable to COIL. 

Meanwhile, TILDE \cite{zhuang2021tilde} follows a query likelihood paradigm, estimating relevance of a passage to a query based on the likelihood of generating the query text from the passage. However, this results in large index sizes as it requires storing inference on the entire BERT vocabulary. TILDEv2 \cite{zhuang2021fastpassagererankingcontextualized} improves upon this by abandoning query likelihood estimation and employing exact term matching. It uses BERT to output a scalar importance weight for tokens only in the passage. In addition, TILDEv2 performs passage expansion using the original TILDE's query token likelihood distribution, significantly speeding up the expansion process compared to auto-regressive query prediction methods like docT5query.

\subsection{Motivation and Our Contributions}
Since the publication of \multibert{} \cite{mallia2021learning}, a number of techniques have been introduced to advance learned sparse retrieval. These improvements include the use of more powerful large language models (LLMs), the idea of filtering queries generated for document expansion \cite{gospodinov2023doc2query}, enhanced training methodologies such as the CoCondenser approach \cite{gao2021unsupervised}, the use of hard negatives \cite{qu2021rocketqa, ren-etal-2021-rocketqav2, zhan2020learning}, and distillation for transferring knowledge learned from powerful cross-encoder models into simpler and more efficient models \cite{formal2022distillation, hofstätter2021improving, lin-etal-2021-batch, santhanam2022colbertv2}. The more recent versions of SPLADE \cite{formal2022distillation,spladev3} have already adopted some of these advances, resulting in impressive effectiveness gains.

In this paper, our objective is to incorporate these cutting-edge techniques into the \multibert{} framework and evaluate the resulting gains in effectiveness and efficiency, along with the associated trade-offs. The enhanced version, which we call \deeperimpact\, achieves significant improvements in retrieval metrics over the earlier version \cite{mallia2021learning}, and substantially closes the effectiveness gap with the recent versions of SPLADE. This is achieved by using Llama 2 for document expansion, initializing with the CoCondenser model, and incorporating hard negatives and distillation during training.

Although the recent versions of SPLADE demonstrate superior retrieval quality, they still suffer from slower processing speeds, while the efficient SPLADE variants lack consistent generalizability across datasets. Therefore, we believe that it is still valuable to explore the efficacy of models like \multibert{} that allow for very fast query processing.

In summary, this paper makes the following contributions:
\begin{itemize}
    \item We propose an optimized version of \multibert{}, which substantially reduces the effectiveness gap with the most effective version of SPLADE without incurring additional efficiency penalties.
    \item We investigate the expansion process and improve it by leveraging a recent and more powerful Llama 2 model.
    \item  We apply effective training strategies using hard-negatives, distillation, and initialization with a pre-trained CoCondenser model.
\end{itemize}

\section{DeeperImpact}

In this section, we provide a detailed explanation of the modifications we attempt in the document expansion strategy and training approach of the original \multibert{} model to enhance its effectiveness.

\subsection{Document Expansion}

This subsection explores methods to improve the document expansion phase, which is crucial for addressing the vocabulary mismatch problem and enhancing the overall retrieval effectiveness. Specifically, we address the challenges associated with hallucinated content in the \doctfquery{} method and consider the adoption of a more advanced language model for more relevant and contextually accurate query generation.

\subsubsection{\textbf{\dtqmm{}}}

As noted by \citet{gospodinov2023doc2query}, the \doctfquery{} expansion technique often has a known inclination to ``hallucinate" non-existent content from the source text, which hampers its retrieval effectiveness while also inflating the index size. Therefore, \dtqmm{} proposes using a relevance model to filter out irrelevant queries prior to indexing---which leads to a 16\% boost in retrieval effectiveness using BM25 scores.

We investigate whether the hallucinations in \doctfquery{} affect \multibert{}'s effectiveness and efficiency. To integrate the \dtqmm{} filtration mechanism into the \multibert{} framework, we first retrieve 80 prospective expansion queries for each document generated using the \doctfquery{} method.\footnote{\url{https://huggingface.co/datasets/macavaney/d2q-msmarco-passage}} These candidate queries are then evaluated using an ELECTRA-based relevance model---mapping each query-document pair to a real-valued relevance score. We utilize the pre-computed relevance scores published by the authors.\footnote{\url{https://huggingface.co/datasets/macavaney/d2q-msmarco-passage-scores-electra}} Specifically, we retain only those queries that achieve a relevance score within the top 30 percentile of all expansion queries across the entire dataset. The choice of top 30\% threshold was based on the findings of the original paper. This selective filtration process ensures that only the most relevant and contextually accurate queries are included in the final expanded document set.

This experiment essentially serves as an evaluation of \multibert{}'s robustness. \multibert---as a sparse retrieval model that learns to assign impact scores to each term in a document---inherently functions as a relevance model. This implies that, ideally, \multibert{} should be capable of suppressing the scores of non-relevant expansion terms introduced as a result of hallucinations from the \doctfquery{} expansion process. Consequently, the integration of the \dtqmm{} filtration mechanism should not lead to substantial improvements in the effectiveness if \multibert{} has been trained correctly. However, should such improvements be observed, this may indicate potential shortcomings in the initial training process or highlight opportunities for further refinement of the model.

\subsubsection{\textbf{Llama 2 Expansions}}

In an effort to enhance the document expansion framework, we experiment with replacing T5 with Llama 2, a more recent and advanced large language model by Meta \cite{touvron2023llama}. Specifically, we fine-tune the pre-trained Llama 2 model with 7B parameters for the downstream task of query prediction.

Unlike T5---which is fully fine-tuned for \doctfquery{}---Llama 2 is a significantly larger model with 7 billion parameters, making full fine-tuning computationally expensive and memory intensive. Therefore we perform the fine-tuning using Low Rank Adaptation (LoRA), which freezes the pre-trained model weights and injects trainable rank decomposition matrices throughout the Transformer architecture \cite{hu2021lora}. We also utilize quantization by loading the weights in 8 bits for further computational and memory efficiency.

We use the same dataset as \doctfquery{} for fine-tuning the pre-trained Llama 2 model i.e. 532k document-query pairs from MS MARCO Passage Qrels Train Dataset \cite{msmarco}. During inference, we use a combined top-k and top-p sampling technique for generating the queries---first selecting the $k$ most likely tokens from the model's vocabulary, then reducing it to the smallest set whose cumulative probability exceeds the threshold $p$, and finally sampling from this reduced set. This balances guiding the model towards probable outcomes while maintaining query diversity.

Additionally, we experiment with the beam search sampling approach. While beam search typically prioritizes queries with higher probabilities and often results in the generation of highly relevant queries, it has a tendency to reduce diversity among the predictions. This approach, while effective in certain scenarios, often limits the breadth of query variations, which is contrary to our aim of enhancing document expansion term diversity to solve the vocabulary mismatch problem.

The overarching objective of modifying the expansion model is to generate more relevant and contextually accurate terms during the document expansion phase. \multibert{} then subsequently learns to score the terms that were not present in the original document. By increasing the quality and relevance of these queries for document expansion, we aim to reduce the vocabulary mismatch problem further, and thereby achieve an improvement in retrieval effectiveness over the original \multibert{} model. 

\subsection{Training Improvements}

We have made several key improvements to enhance our model's training, including the use of a pre-trained CoCondenser checkpoint, the incorporation of hard negatives, and the application of knowledge distillation. Below, we provide a detailed discussion of these improvements.

\subsubsection{\textbf{Pre-Trained CoCondenser Checkpoint}}

Recent advancements in the field of Natural Language Processing (NLP) have incorporated the principles of contrastive learning in order to generate high-quality document representations without the need of manual annotations \cite{gao2021unsupervised, izacard2022unsupervised, wang2021tsdae, wu2020clear}. A prominent approach in Information Retrieval that illustrates this trend is the CoCondenser model \cite{gao2021unsupervised}, which uses the Condenser pre-training architecture \cite{gao2021condenser} in conjunction with a contrastive loss function. This dual approach facilitates the learning of a more refined embedding space from different spans of documents.

The CoCondenser model has demonstrated considerable improvements when used for dense retrieval \cite{gao2021unsupervised}. Specifically, it has shown a significant ability to reduce the sensitivity to noise present in the training data, thereby enhancing stability in the model's performance. Additionally, CoCondenser reduces the reliance on large batch sizes traditionally required for effective learning of the embedding space. Even sparse retrieval models, such as SPLADE, have reported noticeable improvements when initialized with the CoCondenser checkpoint \cite{formal2022distillation}. This is intuitively justified because these models benefit from the high-quality embeddings and robust learning paradigms provided by CoCondenser, which facilitate more accurate information retrieval.

In light of these advantages, we chose to initialize \multibert{} with a pre-trained CoCondenser checkpoint, instead of using the BERT model that is pretrained using Masked Language Modeling (MLM). This choice is motivated by CoCondenser's proven capability to produce superior embeddings through its contrastive learning framework, addressing many limitations associated with MLM-based models.

\subsubsection{\textbf{Hard Negatives}}

One of the primary concerns in training retrieval models is the quality of the negative samples included in the training data. Traditionally, the training of neural rankers, including the original \multibert{} model, has relied heavily on the MS MARCO Triples dataset, which comprises triples of the form $(q,~d^+,~d^-)$, where $q$ is a query, $d^+$ is a relevant (positive) document, and $d^-$ is a non-relevant (negative) document. These negatives are typically sampled using BM25 scores. While BM25 is a robust lexical matching algorithm, it does not fully capture the semantic nuances between relevant and non-relevant documents that are syntactically very similar. Thus, the negatives identified by BM25 often lack the semantic richness that more advanced dense retrieval models are capable of identifying. This limitation can lead to sub-optimal training and subsequently poorer retrieval effectiveness.

We first experimented with the common trick of using multiple in-batch negatives instead of triples. Now, every query $q_i$ in a batch of triples of size $|B|$ will have one positive document $d_i^+$, one negative document $d_i^-$, and a set of $|B| - 1$ additional negative documents $\{d_j^+\}_{j \neq i}$ that are positive documents for other queries in the batch. However, we observed only negligible improvements in the effectiveness of \multibert{} using in-batch negatives.

Addressing the above shortcomings, several recent studies \cite{qu2021rocketqa, ren-etal-2021-rocketqav2, zhan2020learning} have demonstrated the advantages of employing enhanced negative sampling methods to boost the retrieval effectiveness. Enhanced negative sampling methods involve mining \textit{hard negatives}, which are those non-relevant documents that are mistakenly ranked higher by simpler models like BM25. Training with hard negatives compels the model to make finer distinctions between relevant and non-relevant documents, thereby improving its generalization capabilities.

To this end, we make use of the \texttt{msmarco-hard-negatives} dataset\footnote{\url{https://huggingface.co/datasets/sentence-transformers/msmarco-hard-negatives}} published under the Sentence Transformers library \cite{reimers2019sentencebert}. This dataset consists of Top-50 hard negatives for all 800k training queries in the MS MARCO passage dataset. These hard negatives are obtained from BM25 (using ElasticSearch) and 12 different dense retrieval models. This diverse negative sampling strategy introduces a richer set of negatives into the training process. It ensures that \multibert{} does not rely solely on BM25 for negative examples—which may fail to account for nuanced semantic differences—but also leverages hard negatives from dense models, compelling the model to learn deeper semantic distinctions between lexically similar relevant and non-relevant documents, thereby enhancing its overall retrieval accuracy.

\subsubsection{\textbf{Distillation}}

Distillation, a technique for transferring knowledge from a powerful teacher model to a simpler student model, has become a popular method for enhancing the effectiveness of neural rankers in information retrieval tasks. This technique has been extensively validated across both sparse as well as dense retrieval systems \cite{formal2022distillation, hofstätter2021improving, lin-etal-2021-batch, santhanam2022colbertv2}. Although computationally expensive for large-scale retrieval tasks, cross-encoder models, by design, capture more intricate interactions between the query and document, which makes them ideal teacher models for distillation.

In addition to using hard negatives, we integrate knowledge distillation from a cross-encoder teacher model to further enhance \multibert{}'s performance. Again, we make use of the \texttt{msmarco-hard-negatives} dataset which provides not only the Top-50 hard negatives and positives for each training query, but also the relevance scores associated with each pair. These scores are derived from the \texttt{ms-marco-MiniLM-L-6-v2} cross-encoder model.\footnote{\url{https://huggingface.co/cross-encoder/ms-marco-MiniLM-L-6-v2}}

To implement distillation in \multibert{}, we modify the training objective by replacing the original cross-entropy loss function, which considers triples of $(q,~d^+,~d^-)$, with the Kullback-Leibler (KL) divergence loss function. The KL divergence loss involves comparing the score distributions generated by the student model (\multibert{}) and the teacher model (cross-encoder) for each query with its associated positive document, along with the top 50 hard negatives. As such, this loss formulation ensures that \multibert{} learns to approximate the soft relevance score distributions of the teacher model, rather than merely learning from sparse binary labels (i.e., relevant vs. non-relevant).

As an alternative, we also experimented with Margin-MSE loss for distillation, where we optimize the mean squared error of the margin between relevant and non-relevant documents as predicted by the teacher and student models. This approach focuses on relative score differences rather than absolute values. However, our empirical results demonstrate that KL divergence loss yields better performance, leading us to adopt KL divergence loss as the primary distillation method.

\section{Experimental Results}

In this section, we analyze the performance of the proposed changes to \multibert\ using standard test collections and query logs.

\boldparagraph{Hardware.}
The experiments were carried out in memory, utilizing a single thread on a Linux system equipped with dual 2.8 GHz Intel Xeon CPUs and 512 GiB of RAM. For training and running inference on various \deeperimpact\ models, we used 4 NVIDIA RTX8000 GPUs with 48GB of memory each. 

\boldparagraph{Dataset and Query Logs.}
We conduct most of our experiments on the MS MARCO passage ranking \cite{msmarco} dataset.
To evaluate query processing effectiveness and efficiency, we compare with existing methods using the \msmarcodev{}, and we test all methods on the \trecdl{}~\cite{trec2019} and \trecdltw{}~\cite{trec2020} queries from the TREC Deep Learning passage ranking track. Finally, we also provide some metrics for a subset of the BEIR datasets.

\boldparagraph{Baselines.}
In our experiments, we compare our proposed \deeperimpact\ model against \unicoil\ with TILDE document expansion, \splade\ (which refers to {CoCondenser-EnsembleDistil} in \cite{formal2022distillation}), the recent SPLADE-v3 \cite{spladev3} and the original \multibert\/  model.

\boldparagraph{Implementations.}
We use Anserini~\cite{anserini} to generate the inverted indexes of the collections. We then export the Anserini indexes using the CIFF common index file format \cite{ciff}, and process them with PISA~\cite{pisa} using the MaxScore query processing algorithm~\cite{maxscore}. 
For \unicoil, \splade\, and \multibert, we use the indexes made available in Anserini. Since SPLADE-v3 is fairly recent, we only report the scores that are present in their paper.

Similar to the original \multibert{} paper, we also employ quantization for storing real-valued document-term impact scores in the inverted index. We adopt an 8-bit linear quantization strategy, which preserves precision without significantly increasing space requirements. This allows us to efficiently compute query-document scores by summing the quantized scores of the matching document terms.

Both training and inference tasks of \deeperimpact\ are implemented in Python using the PyTorch library. The training for \deeperimpact\ was carried out using a maximum length of 300 input tokens, while for inference, a maximum token length of 512 was used.

We downloaded the pre-trained Llama 2 7B model weights from Meta's website and converted them to the HuggingFace format. We perform the fine-tuning in Python with the \texttt{peft} library using Low Rank Adaptation (LoRA) with $r = 8$, along with 8-bit quantization. After fine-tuning, we merge the adapter weights and sample 80 queries for each document in the MS MARCO passage dataset using top-k $(k = 50)$ sampling and top-p $(p=0.95)$ sampling.

Our source code is publicly available.\footnote{\url{https://github.com/basnetsoyuj/improving-learned-index}} We will also publish the expansions generated by Llama 2 for the MS MARCO passage collections.

\boldparagraph{Metrics.}
To measure effectiveness, we use mean reciprocal rank (MRR@10), normalised discounted cumulative gain (NDCG@10), and recall (R@1k). We also report recall at different cutoff values. Finally, we compute the mean response time (MRT) for every model in ms.

\subsection{Overall comparison}
Our first experiment aims to compare the retrieval effectiveness of the proposed \deeperimpact\ model with the original \multibert{}, \unicoil\ with TILDE document expansion, \splade\ (which refers to {CoCondenser-EnsembleDistil} in \cite{formal2022distillation}), and SPLADE-v3. We also report the retrieval effectiveness of \multibert\ when combined with \dtqmm{}, which we will refer to as \multibert{}{-}\textit{-}. The results are presented in Table~\ref{tab:overall}, which shows effectiveness and efficiency for the three query logs on MS MARCO. We retrieve the top 1000 documents for each query, without re-ranking, and report the values of NDCG@10, MRR@10, and Recall@1k, as well as MRT.

We first notice that integrating \dtqmm{} filtration into the \multibert{} framework does not substantially improve its effectiveness and in some cases results in marginal declines. This strengthens our claim that \multibert{} is already trained correctly as a relevance/filtration model. Instead of filtering out the predicted queries, it makes more sense to provide \multibert{} with a larger set of expanded documents regardless of the fact that they might contain terms predicted as a result of hallucination.

We also notice that \deeperimpact\ demonstrates substantial improvements on the \msmarcodev{}, \trecdl{}, and \trecdltw{} queries compared to the original \multibert{} model. Specifically for the \msmarcodev{}, \deeperimpact\ achieves an NDCG@10 score of 43.49, which is notably higher than the 38.65 of \multibert. Additionally, \deeperimpact\ shows better effectiveness in MRR@10 (37.30 versus 32.74) and Recall@1k (96.83 versus 94.76). These improvements can be attributed to the better expansion strategy with Llama 2, incorporation of hard negatives and distillation during training, and the use of the CoCondenser checkpoint---which we will look at in detail in the subsequent sections.

It is worth noting that while \unicoil\ performs worse both in terms of effectiveness and efficiency on the \msmarcodev{}, it achieves better effectiveness on the \trecdl{} and \trecdltw{} queries, surpassing \deeperimpact{} in MRR@10 and Recall@1k. Additionally, it has a much lower MRT for both \trecdl{} and \trecdltw{} queries. This suggests that \unicoil\ may be more effective for certain types of queries within these datasets.

SPLADE consistently performs better in terms of the MRR@10, Recall@1k, and NDCG@10 metrics compared to the \deeperimpact{} model. However, while \splade\ remains dominant in terms of retrieval effectiveness, it is substantially less efficient than \deeperimpact, with an MRT of 378.00 ms compared to 39.62 ms for \deeperimpact\ on the \msmarcodev{}. This trend is also consistent across the \trecdl{} and \trecdltw{} queries.

These results underscore the retrieval effectiveness of the \deeperimpact\ model while continuing to maintain high retrieval efficiency. This balance is particularly advantageous for large-scale information retrieval tasks, where computational efficiency and response time are critical. Given these strengths, \deeperimpact\ presents a compelling alternative to models like \splade\ for scenarios where both effectiveness and efficiency are important. However, if the primary focus is solely on retrieval effectiveness, \splade\ may still be preferred.

\begin{table}[t]
\caption{Effectiveness metrics and mean response time (MRT, in ms) for first-stage methods, on \msmarcodev, \trecdl queries, and \trecdltw queries. For the recent SPLADE-v3, only the scores available in their paper have been reported.}

\centering
\begin{tabular}{lrrrr}
\toprule
 & \multicolumn{1}{r}{NDCG@10} & \multicolumn{1}{r}{MRR@10} & \multicolumn{1}{r}{R@1k} & \multicolumn{1}{r}{MRT} \\ 
\midrule
\multicolumn{5}{c}{\msmarcodev} \\
\midrule
\unicoil & 41.19 & 34.96& 96.46 & 49.18\\
\splade & 44.98 & 38.38 & 98.33 &  476.65\\
SPLADE-v3 & - & 40.20 & - & - \\
\multibert  & 38.65        & 32.74        & 94.76  & 33.70\\
\multibert{-}\textit{-} & 37.63 & 31.75 & 92.58 & 21.73\\
\deeperimpact & 43.49 & 37.30 & 96.83 & 39.62 \\
\midrule
\multicolumn{5}{c}{\trecdl}  \\
\midrule
\unicoil & 70.69 & 98.84 & 83.15 & 53.42\\
\splade & 73.23 & 97.29 & 87.35 & 468.56\\
SPLADE-v3 & 72.30 & - & - & - \\
\multibert  & 65.85 & 89.53 & 75.46 & 170.44 \\
\multibert{-}\textit{-} & 67.13 & 93.80 & 73.37 & 118.44 \\
\deeperimpact & 67.88 & 89.73& 82.22 & 145.93 \\
\midrule
\multicolumn{5}{c}{\trecdltw}  \\
\midrule
\unicoil &68.51 & 94.75 & 83.91 & 60.03 \\
\splade &71.87 &95.52 & 89.91 & 525.56\\
SPLADE-v3 & 75.40 & - & - & - \\
\multibert  & 60.38 & 87.07 & 78.87 &  188.00 \\
\multibert{-}\textit{-} & 59.65 & 87.47 & 76.39 & 124.29 \\
\deeperimpact  & 68.54 & 92.13 & 80.69& 169.37\\
\bottomrule
\end{tabular}
\label{tab:overall}

\end{table}

\subsection{Ablation Study}

We also conduct an ablation study to rigorously evaluate the contributions of various components in enhancing the effectiveness of our proposed \deeperimpact\ model. We begin with a simplified version of \multibert\ without any document expansions and trained using the MS MARCO Triples, referred to as \deeperimpact{}$_{Base}$. Subsequently, we augment this base model with the following series of improvements:

\begin{enumerate}
    \item \textbf{+ Llama 2 expansions}: Incorporates document expansions generated by the fine-tuned Llama 2 model.
    \item \textbf{+ CoCondenser}: Substitutes the pre-trained \texttt{BERT-base-uncased} backbone with the CoCondenser checkpoint.
    \item \textbf{+ Distillation}: Trains the CoCondenser-enhanced model with hard negatives and distillation from a cross-encoder (using KL divergence loss).
\end{enumerate}

The results are summarized in Table~\ref{tab:ablation}. Our performance metrics include MRR@10, Recall@1k, and MRT for the \msmarcodev{}. We notice incremental improvements in the MRR@10 as we continue adding the refinements. The most substantial improvement in MRR@10 and Recall@1k is observed when incorporating Llama2 expansions, which can be attributed to its effectiveness in addressing vocabulary mismatch issues. We can further compare these metrics to those reported using other models like DocT5Query (\multibert{}) and \dtqmm{} (\multibert{-}\textit{-}) in Table~\ref{tab:overall}, and observe that using Llama2 expansions demonstrate better retrieval effectiveness in both metrics.

Interestingly, subsequent enhancements with CoCondenser and distillation do not further improve recall, suggesting that even Llama2 expansions trained with the triples are as effective in capturing the relevant documents. However, these enhancements do contribute to better MRR@10, indicating improved ranking precision. This implies that while document expansion substantially aids in ensuring a broader recall of relevant documents, the fine-tuning with CoCondenser and distillation techniques improves the ranking, likely due to better contextualized representations and more robust handling of difficult queries.

Notably, the base model without expansions exhibits a much lower MRT, which is expected due to its smaller postings list. The expansions inflate the index size, leading to increased MRT. 

\begin{table}[t]
\caption{Ablation study on \msmarcodev evaluating effectiveness metrics and mean response time (MRT, in ms) as components are incrementally added to \multibert{}.}
\centering
\begin{tabular}{lcccc}
\toprule
Strategy &  & \multicolumn{1}{c}{MRR@10} & \multicolumn{1}{c}{Recall@1000} & \multicolumn{1}{c}{MRT} \\ 
\midrule
\multicolumn{5}{c}{\msmarcodev} \\
\midrule
\rowgroup{\deeperimpact$_{Base}$} &&  27.32 &  88.07 & 8.93\\
+ Llama2 expansions && 33.42 & 96.83 & 43.25 \\
+ CoCondenser  && 35.36 & 96.75 & 42.38\\
+ Distillation && 37.30   &     96.83 & 39.62\\
\bottomrule
\end{tabular}
\label{tab:ablation}
\end{table}

\subsection{\doctfquery{} vs. Llama 2 Expansions}
In this section, we compare the quality of document expansions generated by \doctfquery{} and Llama 2 using BM25 scores on the \msmarcodev{} dataset. The effectiveness metrics are summarized in Table~\ref{tab:document-expansion-comparison} for varying numbers of queries appended to each document in the corpus.

From the results, we observe that the MRR@10 scores for both \doctfquery{} and Llama 2 expansions are quite similar across different numbers of queries appended to each document. This similarity suggests that any improvements in the MRR of the \deeperimpact\ model are primarily due to the superior architecture (i.e., CoCondenser) and the advanced training objective (i.e., hard negatives and distillation) rather than the quality of the expansions themselves.

However, Llama 2 expansions consistently outperform \doctfquery{} expansions in terms of the recall metrics. This indicates that Llama 2 expansions are more effective at retrieving a broader set of relevant documents, as they address vocabulary mismatch better which enhances recall.


\begin{table}[t]
\caption{Comparison of effectiveness metrics (MRR@10 and Recall@1000) between \doctfquery{} and Llama 2 expansions for varying numbers of appended queries using BM25 scores on \msmarcodev .}
\centering
\begin{tabular}{lcccccc}
\toprule
 & \multicolumn{2}{c}{MRR@10} & & \multicolumn{2}{c}{Recall@1000} \\
\cmidrule{2-3} \cmidrule{5-6}
n & \multicolumn{1}{c}{\doctfquery{}} & \multicolumn{1}{c}{Llama 2} & & \multicolumn{1}{c}{\doctfquery{}} & \multicolumn{1}{c}{Llama 2} \\
\midrule
5  & 25.9 & 25.4 & & 92.9 & 94.0 \\
10 & 26.5 & 26.5 & & 93.9 & 94.8 \\
20 & 27.2 & 25.4 & & 94.4 & 95.4 \\
40 & 27.7 & 27.4 & & 94.7 & 95.5 \\
80 & 27.8 & 27.4 & & 94.5 & 95.6 \\
\bottomrule
\end{tabular}
\label{tab:document-expansion-comparison}
\end{table}

\subsection{Recall at Different Depths}

Table~\ref{tab:recall-at-diff-depths} presents the recall metrics at various depths on the \msmarcodev dataset for four different models: uniCOIL (with TILDE document expansion), \splade\ (which refers to {CoCondenser-EnsembleDistil} in \cite{formal2022distillation}), \multibert\, and \deeperimpact. We notice that SPLADE consistently achieves the highest recall at all the depths. However, \deeperimpact---which offers much faster query processing---also demonstrates strong effectiveness, keeping ahead of uniCOIL in all the depths.

Remarkably, \deeperimpact\ attains a recall at depth 400 that is equivalent to the Recall@1000 achieved by the original \multibert\ model. This highlights the effectiveness of \deeperimpact\ in retrieving relevant documents at shallower depths, demonstrating its efficiency in reducing the number of retrieved documents necessary to achieve high recall.

\begin{table}[t]
\caption{Comparative analysis of recall at various depths on \msmarcodev for \unicoil{}, \splade{}, \multibert{}, and \deeperimpact{}.}
\centering
\begin{tabular}{lcccc}
\toprule
 Recall & \multicolumn{1}{c}{\unicoil}  & \multicolumn{1}{c}{\splade} & \multicolumn{1}{c}{\multibert} & \multicolumn{1}{c}{\deeperimpact}  \\ 
\midrule
10  &    62.00     & 67.07 & 58.59 &  64.41 \\
100 &    86.78     & 90.96 & 84.21 & 87.64\\
200 &    90.72     & 94.72 & 88.25 & 91.51\\
400 &  93.99 & 96.87 & 91.84 & 94.56\\
600 &  95.21      & 97.61 & 93.38 & 95.75 \\
800 &   96.06     & 98.09 & 94.04 & 96.36\\
1000  &  96.46    & 98.33 & 94.79 & 96.83 \\
\bottomrule
\end{tabular}
\label{tab:recall-at-diff-depths}
\end{table}

\subsection{BEIR Zero-Shot Evaluations}

The BEIR (Benchmarking IR) benchmark is a comprehensive evaluation suite that encompasses a wide variety of Information Retrieval (IR) datasets designed to rigorously test the zero-shot effectiveness of retrieval models. To comprehensively evaluate the effectiveness of \deeperimpact\, we benchmarked it against SPLADEv3-doc, SPLADEv3, and the large efficient SPLADE variant BT-SPLADE-L \cite{Lassance_2022} on a subset of BEIR datasets using NDCG@10 as the metric. The zero-shot inferences were performed with \deeperimpact\ model trained on MS MARCO passage dataset. We also used the Llama 2 model---fine-tuned for query prediction on MS MARCO passage dataset---for expanding each of these datasets with relevant queries. The results are displayed in Table~\ref{tab:beir}.

Among the SPLADE variants, SPLADE-v3 consistently achieves the highest NDCG@10 scores across the datasets, demonstrating its state-of-the-art effectiveness. However, it is worth noting that SPLADEv3 incorporates query processing, making direct comparisons less straightforward.

SPLADE-v3-doc, which processes queries as a bag of words without additional query processing, is a more appropriate baseline for comparison with \deeperimpact{}. In this context, \deeperimpact\ exhibits competitive effectiveness, often closely aligning with SPLADEv3-doc. Its effectiveness is also fairly close with the large efficient SPLADE variant. 

In summary, this BEIR evaluation underscores the strength of \deeperimpact\ in competing with advanced retrieval models like SPLADE, especially when evaluated under comparable conditions.

\begin{table}[t]
\caption{NDCG@10 results across a subset of BEIR datasets for \deeperimpact{}, BT-SPLADE-L, SPLADE-v3-doc, and SPLADE-v3.}
\centering
\begin{tabular}{lcccc}
\toprule
 Dataset & \multicolumn{1}{c}{\deeperimpact} & \multicolumn{1}{c}{BT-SPLADE-L} & \multicolumn{1}{c}{SPLADE-v3-doc} & \multicolumn{1}{c}{SPLADE-v3}  \\ 
\midrule
TREC-COVID  & 70.3    & 66.06     & 68.1          & 74.8      \\
Scidocs     & 15.2    &  15.25   & 15.2          & 15.8      \\
NFCorpus    & 33.2    &  33.1   & 33.8          & 35.7      \\
Touché-2020 & 26.9    & 27.03     & 27.0          & 29.3      \\
FiQA-2018   & 30.8    & 31.77     & 33.6          & 37.4      \\
SciFact     & 65.4    & 67.43   & 68.8          & 71.0      \\
Quora       & 72.8    & 72.34    & 77.5          & 81.4      \\
ArguAna     & 34.9    & 47.29     & 46.7          & 50.9      \\
\bottomrule
\end{tabular}
\label{tab:beir}
\end{table}

\section{Conclusions and Future Work}

In this paper, we presented \deeperimpact\, an optimized version of the \multibert{} sparse retrieval model, aimed at improving retrieval effectiveness while maintaining computational efficiency. By substituting \doctfquery{} with a fine-tuned Llama 2 model for document expansion, leveraging advanced training strategies such as initializing with CoCondenser, employing hard negatives, and integrating distillation techniques, \deeperimpact\ substantially narrows the effectiveness gap with state-of-the-art but more expensive sparse retrieval models such as SPLADE. Future work could focus on further enhancing the training of \deeperimpact\ using a combination of training loss objectives and searching for optimal hyperparameters similar to what has been done with recent versions of SPLADE. We would also like to experiment with better ways to perform document expansion, as it is currently the computational bottleneck before training \multibert{} and performing inferences on a new dataset. Ideally, we would like to transition from auto-regressively predicting queries to sampling relevant query terms from a query likelihood distribution---drawing inspiration from the TILDE expansion algorithm. This new approach would not only allow for a greater degree of unique term expansion but also offer significantly enhanced efficiency, thereby further mitigating the issue of vocabulary mismatch.

\clearpage
\bibliography{sample-1col}

\end{document}